%% file: main.tex
\documentclass[a4paper]{article}
\usepackage{enumitem}
\usepackage{calc}
\usepackage{anysize}
\marginsize{2cm}{2cm}{1cm}{2cm}
\usepackage{fancyhdr}
\usepackage{cite}

\usepackage{authblk}
\usepackage{soul}
\usepackage[dvipsnames]{xcolor}
\usepackage{todonotes}
\usepackage{placeins}
\usepackage{graphicx}
\usepackage{hyperref}
\hypersetup{
    colorlinks=true,
    linkcolor=black,
    citecolor=CadetBlue,
    filecolor=CadetBlue,
    urlcolor=CadetBlue,
}
\usepackage{xurl}
\usepackage{lipsum}
\usepackage{multicol}
\usepackage{setspace}
\usepackage[version=4]{mhchem}  

\fancypagestyle{plain}{%
  \fancyhf{}
\rhead{HECAP.ECO/002(2025)}
}

\renewenvironment{abstract}
 {\par\noindent\textbf{\abstractname}\ \ignorespaces \\}
 {\par\noindent\medskip}

\providecommand{\keywords}[1]
{
	\small	
	\textbf{\textit{Keywords---}} #1
}

\title{\huge{ORLCA: A concept for an open-source Life Cycle Assessment repository built for research}}
\author[1]{Hannah Wakeling\thanks{hannah.wakeling@physics.ox.ac.uk}}
\author[2]{Kristin Lohwasser}
\author[3]{Peter Millington}
\affil[1]{\small John Adams Institute for Accelerator Science at University of Oxford, Denys Wilkinson Building, Keble Road, Oxford, OX1 3RH, UK}
\affil[2]{\small School of Mathematical and Physical Sciences, University of Sheffield, Hicks Building, Hounsfield Road, Sheffield, S3 7RH, UK}
\affil[3]{\small Department of Physics and Astronomy, University of Manchester, Manchester M13 9PL, UK}

\begin{document}
\pagestyle{fancy} 
\fancyhead[R]{HECAP.ECO/002(2025)}
\pagenumbering{arabic}
\renewcommand*{\thefootnote}{\fnsymbol{footnote}}
\maketitle

\begin{figure}[h!]
    \centering
    \includegraphics[width=0.45\linewidth]{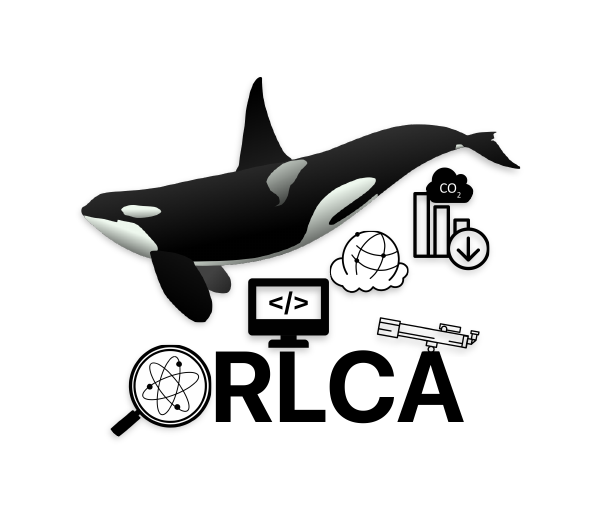}
    \label{fig:placeholder}
\end{figure}
\begin{center}
Please consider reading this document via the least environmentally impactful method available to you.
\end{center}

{\color{gray}\hrule}
\vspace{0.4cm}
\begin{abstract}

\input{./abstract.tex}
\end{abstract}
{\color{gray}\hrule}

\newpage
\section{Introduction}


There is a scientific consensus that climate change is occurring now, with unequivocal anthropogenic influence~\cite{IPCC_2021, ScientificConsensus}. 
There have been multiple calls for action from scientists from various specialities~\cite{Overpeck_climate, Shrivastava_Kasuga_Grant_2023, Hagedorn, HECAPplus, Boisvert, A4E, ScientistsForFuture, XR_scientists}. One of these actions is to actively work on reducing the environmental impact of research.

Carrying out a Life Cycle Assessment (LCA) is often the best way to identify the dominant negative environmental externalities and to determine mitigations that can be employed to reduce them. 
Within research, LCAs help researchers and engineers to design, construct, operate, and decommission their facilities more sustainably. 
Indeed, large-scale facilities and Research Infrastructures (RIs) such as those within the fields of particle physics, accelerator physics and astronomy are starting to embed sustainability into their design and operation to understand how their environmental impacts are distributed by performing LCAs~\cite{GRANDlca,ruedi,AthenaTele,ARUPstudyCLICILC}.
The process of performing LCAs within these facilities has, however, revealed the many challenges in undertaking them in these fields.
LCAs require detailed accounting of every resource consumed throughout a product's or system's lifecycle, making the sheer scale of some RIs exceptionally complex to assess. Other facilities, particularly smaller projects, struggle to perform LCAs due to data availability, financial costs, time constraints, and availability of expertise. Whilst there have been recent LCA efforts within these fields, data accessibility and availability, and access to appropriate training remain common and time-consuming obstacles for all types of RIs and facilities in these disciplines. This is often a significant impediment in environmental efforts, further detailed in section~\ref{ref:LCAstruggles}. A way to remedy this is the establishment of a repository that seeks to facilitate environmental impact assessments and LCAs within research by centralising tailored LCA data in an open-access LCA repository. To this end, this paper proposes the concept of the Open Research LCA (ORLCA) repository. The ORLCA repository (pronounced ``orca'') is designed to aid in the advancement of environmental impact knowledge for research by providing free, open-access, and time-saving resources, see Section~\ref{ref:ORLCA}, with benefits and limitations described in Sections~\ref{sec:benefits} and~\ref{sec:limits}. ORLCA has already garnered significant support within the community and is currently seeking funding for personnel and implementation. Researchers who are interested in sustainability in research are encouraged to engage with ORLCA, whether through advocating for its need, or contributing time, expertise, or data. In addition, ORLCA welcomes contact from groups or institutions who are aware of collaborative opportunities or potential funding avenues.

\section{Life Cycle Assessments}\label{ref:LCA}

Life Cycle Assessments (LCAs) are systematic methods for evaluating the environmental impacts associated with each lifecycle stage of a product or system, from raw material extraction and manufacturing to use, disposal, and recycling. 
They have been standardised in ISO 14040~\cite{ISO14040} and ISO 14044~\cite{ISO14044}. 
LCAs are a well-regarded and accepted method to evaluate a range of environmental impact factors associated with products or services throughout their whole lifecycle. 
They can be used as research tools within emerging technologies, for design comparisons, hotspot evaluation, and to inform policy. 

For example, LCAs can be used to pinpoint the most significant environmental impacts in each stage of a product, from raw material extraction to end-of-life disposal.
They require an inventory of each and every resource used in the construction, operation and decommissioning of a product or system, depending on the scope of the LCA. 

By providing a holistic picture rather than focusing on a single phase or process, LCAs help analysts identify hidden trade-offs and opportunities for environmental impact improvement. In research and large-scale scientific endeavours, LCAs are particularly relevant because they provide a framework to assess the sustainability of complex infrastructures, technologies, and practices. 
From laboratory facilities to particle accelerators to data centres, LCAs enable researchers and policymakers to quantify environmental concerns, compare alternative designs, and guide decision-making toward more sustainable innovations. 
This makes LCAs ideal tools for environmental accounting and for aligning research with global sustainability goals.

In today's LCA database landscape, there exist a few flagship, comprehensive, mature, proprietary packages for general products and sectors. Their pricing depends on the license class and version, but can often reach thousands of Euros (EUR) per user. There are also many high-quality open-access or open-source databases which have coverage of open or sector-specific domains. These databases are all indispensable for modelling the bulk environmental impacts of RIs and facilities -- especially concrete, basic metals and electricity use -- however, as will be discussed in Section \ref{ref:LCAstruggles}, these databases lack detailed inventories for many components that are central to RIs.


\section{Challenges of life cycle assessment in a research context}\label{ref:LCAstruggles}

RIs, facilities and smaller-scale laboratories or experiments, often require the purchase of specialised components, production of bespoke components, use of uncommon resources, creation of new materials, chemicals or processes, and sometimes disposal of radioactive materials. 
These LCA data are often unavailable for multiple reasons; they can:
\begin{itemize}
    \item be inaccessible due to being proprietary, unpublished, or behind a paywall.
    \item be buried in published papers (sometimes also behind a paywall).
    \item be out of date or geographically ill-suited.
    \item have insufficient granularity or accuracy of processes.
    \item simply not exist as the bespoke components, materials, or processes have never been evaluated.
\end{itemize}
Whilst some recent LCA commissions for RIs (the majority of which were carried out by commercial companies at financial cost) and subsequent publications mark a significant step towards embedding environmental sustainability into these research fields, data accessibility and availability remain common and time-consuming obstacles.
Other projects, particularly smaller projects, also struggle to perform LCAs due to financial costs, time constraints, availability of expertise, or access to appropriate training.
Whilst there have been recent LCA efforts within these fields, these issues remain obstacles for all types of facilities in these disciplines.
This is often a significant impediment to environmental efforts.

As a case study, a proposed multi-stage mega-project---the ISIS-II Neutron and Muon Source~\cite{ISIS-II}---encountered many barriers in performing an LCA during the feasibility and design phase using open-source LCA databases. 
Often, materials, chemicals, components, machining techniques, and construction techniques were either not available, considered too out-of-date to be acceptable models, were not regionally representative, or not accurate representatives for the ISIS-II LCA. 
The practitioner had to rely on the creation of proxies, engineering concepts, bills of materials, and other less accurate modelling alternatives.
To improve access to materials and a few very common, general components, the choice was made to purchase a license for the commercial ecoinvent LCA database. 
This enabled the LCA process; however, no existing database, paid-for or otherwise, was able to provide the data required for a complete LCA of all components and materials of the proposed facility.

The concept of an open LCA repository for research would enable researchers and facilities, particularly those without dedicated environmental sustainability funding, to assess and minimise their environmental impacts.

\section{The ORLCA Concept}\label{ref:ORLCA}

The concept for the ORLCA repository is to develop a first-of-its-kind open-source, curated repository of LCA inventory specifically for research, but available for use by others beyond. 
This proposed repository would consist of collated general and bespoke LCA data. 
The aim is that the ORLCA repository facilitate comprehensive LCAs, enable sustainable design choices, and ultimately contribute to environmentally responsible, cutting-edge research by providing unrestricted and sustainable access to data and solving the issues of data accessibility, availability, financial costs, and expertise. 
In addition, by creating a single and openly accessible repository, LCAs can be cross-checked and compared fairly. 
To determine the choice of open-source license, consideration of the best options will be taken during the reconnaissance stages of the project.
As a consequence of the ORLCA repository being open-source, existing software, databases and repositories would be able to access and use the ORLCA repository, expanding its reach.

The ORLCA concept aims to provide new data products in the form of research life cycle inventories, materials and LCA results (simulated and measured).
The ORLCA concept is proposed to develop a repository of these data, which can then be imported into a range of LCA software (with priority of open source or free-to-access software) like OpenLCA~\cite{openlca}, through the use of a common LCA format (e.g., JSON-LD or ILCD).

The general-purpose open repository Zenodo~\cite{https://doi.org/10.25495/7gxk-rd71} is an ideal platform to host ORLCA. Developed under the European OpenAIRE program, it is operated by CERN, a large particle physics research infrastructure. The platform hosts digital data records, such as data sets, research software and reports. It also allows creation of shared areas, so-called ``communities", that can be managed by a team, for example for conferences or specific subfields of a domain (e.g., HEP data for Machine Learning). Data records are associated with a community, making them easier to find and use coherently. 

The ORLCA concept would need to ensure completeness and stand-alone use of the repository and ensure all relevant data supplements the contributions from users. 
Stand-alone use is important to enable benchmarking between studies. 
In the initial stages of the database, existing open-source databases will be utilised to provide the baseline of the ORLCA database of bulk materials and generic processes.
The ORLCA repository is proposed to be expanded and populated with LCA data from partners and third-party data from the relevant fields, relevant third-party data from intersectional fields, and relevant third-party materials data. 
Further investigation into the methods of data collection will be an essential reconnaissance stage of the ORLCA project. 
The wide range of data locations and formats, and access to existing data will require a multi-faceted approach for data compilation. 
In addition, data collection will be required where it does not currently exist. 
Initial ideas include the use of methods of data submission, ethical data scraping, data digitisation, and experimental data collection (including via the Internet of Things~\cite{IoT}), including responsible use of machine learning and artificial intelligence to do so~\cite{LCA-automation}.

Users would be able to utilise the ORLCA repository and contribute their own domain-specific LCA data, which has the additional benefits of facilitating reproducibility of results and preventing duplication of efforts. 
Where detailed inventory are not available, high-level data with uncertainties would be provided as a default. 
The managers of the ORLCA repository would identify the necessary data tailored to the specific needs of these fields, collate available open data from partners and third-party sources, and gather information on critical gaps within the data (and attempt to fill them). A quality standard document will be formulated to describe the formats to be followed and the detail levels of data and documentation as well as maximum uncertainties on the data that are required for data to be accepted. These standards should be set in collaboration with the users of the repository.
A community contribution framework with tutorials (including on how to collect and calculate LCA data), data submission guidelines, a peer-review process, and a citation/DOI assignment would be required for the contributed datasets. 
Legal compliance would need to be ensured by providing attribution for data and assigning the appropriate licenses to the contributed data. 
Roles define who can review, accept or decline submissions, edit records metadata and generally manage the community settings, allowing for quality control, e.g., to ensure adherence to a common file format. Standards and expectations will be defined in community standards and a public Curation Policy. Each data record is associated with a persistent digital object identifier (DOI) and can be updated, creating a new unique DOI. This allows for unambiguous versioning and citing, including of training materials. Further control over additional software to support the usage of ORLCA can be maintained by linking to GitHub~\cite{github}. ORLCA would thus be designed for sustained and ongoing use and contributions beyond any financial or temporal boundaries, in theory for as long as the Zenodo database and the GitHub software repository are available to the public. The ORLCA repository should remain adaptable to future collaboration and expansion into additional research disciplines.

In summary, the aim of the ORLCA repository is to consolidate existing but currently disjointed efforts in environmental impact assessments across the fields. This will require crowd-sourcing of information from members across all disciplines including, e.g., contribution from RIs and facilities that have performed LCAs or collected inventory data. In turn, the ORLCA concept would foster informed decision-making across product lifecycles, encompassing material selection, energy consumption, waste management, and other sustainability practices.



\section{Scientific benefits}~\label{sec:benefits} 

The ORLCA repository would allow researchers to assess material and operational impacts without having to start an LCA from scratch. 
This could save months or even years of iterative sustainable design and would have the additional benefit of not requiring the researchers to become experts in sustainability in their research areas to perform these studies, accelerating the research on the environmental impacts of components and new technologies.
Significant cost-savings (e.g., by reducing resource consumption and LCA consultancy costs) could lead to reinvestment of these savings towards scientific endeavours. 

The ORLCA repository could also facilitate interdisciplinary collaboration and innovation in eco-design by providing open-access LCA data as a common language between these communities. In addition, integrating open LCA data from the ORLCA repository into digital twins~\cite{DigiTwin} of facilities and component designs offers a novel approach to evaluating and facilitating sustainable design decisions prior to physical prototyping.
This also stimulates novel research directions, including, e.g., the comparison of different types of accelerator and component design with an environmental sustainability lens.  Therefore, the ORLCA repository would promote sustainable design and environmentally responsible scientific research by informing decisions on areas such as production material choices, procurement, operations, and decommissioning.

The facilitation of consistent and comparable reporting of environmental impacts across different designs using the same framework enables benchmarking in the form of product comparisons and progress of a facility’s environmental sustainability over time. Due to the diverse methodologies and databases employed for LCA across the different fields, RIs and facilities, benchmarking has not yet been feasible.

Finally, the potential LCA outputs resulting from the use of the ORLCA repository would facilitate users in the preparation of environmental impact statements within funding applications, and further inform funding and science agencies on the environmental impacts of research to be able to design more environmentally focused funding frameworks.

\section{Scope and limitations}\label{sec:limits} 

The fields of particle physics, accelerator physics and astronomy are proposed as a first target for the ORLCA repository development, given their extensive reliance on big science infrastructure, their growing emphasis on sustainability, and the challenges already identified for them in accessing relevant LCA data. The ORLCA repository will nevertheless be designed with sufficient flexibility and modularity that data relevant for other fields and subdisciplines can readily be incorporated.
Additionally, these fields are currently engaged in strategic community planning and proposal development for the forthcoming decades of research (including the European Strategy for Particle Physics Update (ESPPU)~\cite{ESPPU}, SNOWMASS~\cite{SNOWMASS}, and Science Vision for European Astronomy~\cite{ScienceVisionAstro}). 
Notably, these fields and their strategy documents have (and should continue to have) increasing integration of environmental sustainability in their remit. This timeliness offers an important opportunity to promote and grow support for the ORLCA concept within these fields.

One of the largest open questions regarding the proposed ORLCA concept is the question of longevity and self-sustainability of the concept.
Once the initial structure for the repository exists, it would be possible for the repository to continue to be updated and expanded with third-party and crowd-sourced data with minimal upkeep and personnel requirements. 
As an open resource on Zenodo and GitHub (at a minium), community contribution to the original ORLCA database would be possible with minimal review. 
The setup would be designed to ensure simplicity of this process, such that reviewing is as simple as possible.

Through successful deployment, advertisement, popularisation, and easy use of ORLCA for all stakeholders, researchers would, after collecting LCA data -- from compiling a bill of materials and evaluating the power consumption of a component, to performing a full environmental impact assessment -- be empowered to share their data through the ORLCA repository.

\section{Conclusion}

Open access to LCA inventory, process and product data would be a crucial enabler for LCAs in research. 
By providing open access to, and training for, a domain-specific LCA repository, the ORLCA repository would enable LCAs where they would otherwise not be feasible and significantly speed up their execution. 
The platform would promote consistent environmental impact reporting and benchmarking, support long-term sustainability progress tracking, and prevent duplication of efforts. 
ORLCA could also encourage interdisciplinary collaboration and new eco-design research, and inform funding agencies on infrastructure environmental impacts.
The ORLCA concept offers an important opportunity for research to help shape the landscape of environmental impact assessment.

\section{Acknowledgements}

The work of HW was supported by the Science and Technology Facilities Council (STFC) via the John Adams Institute (JAI), University of Oxford [Grant No. ST/V001655/1].
The work of PM was supported by the Science and Technology Facilities Council (STFC) [Grant No.\ ST/X00077X/1] and a United Kingdom Research and Innovation (UKRI) Future Leaders Fellowship [Grant No.\ MR/V021974/2].
This work is supported by the \href{https://www.hecap.eco/}{Sustainable HECAP+ Initiative}, which acknowledges funding and in-kind support from STFC Environmental Sustainability. 
For their strong support of the ORLCA concept, thanks to:
the Accelerator Science and Technology Centre (ASTeC), STFC; 
the French National Centre for Scientific Research (CNRS) Nucléaire et Particules; 
the European Particle Physics Laboratory Directors Group (LDG) Working Group on the Sustainability Assessment of Accelerators;
the European Plasma Research Accelerator with Excellence in Applications (EuPRAXIA) Consortium Collaboration; 
the Extreme Light Infrastructure (ELI) European Research Infrastructure Consortium (ERIC); 
the ISIS Neutron and Muon Source, STFC; 
the University of Manchester Department of Physics; 
the Durham University Department of Physics contribution to Multi-Object Spectrograph (MOSAIC); 
the University of Sheffield School of Mathematical and Physical Sciences; 
and UKRI Environmental Sustainability.
Thanks to Michael D\"uren, Mandeep Gill, Diego Herrera Ruiz, Rakhi Mahbubani, and Ruth Poettgen for their advice and aid.
Finally, the temporary ORLCA logo uses public domain images from Wikimedia commons. Costs for professional logo design are incorporated in ORLCA funding applications.

\bibliographystyle{IEEEtran-hmw}
\bibliography{references}
\end{document}

%% file: abstract.tex
Comprehensive Life Cycle Assessment (LCA) as a tool to account for the full range of environmental impacts of resource use in commodities or services is a first step in reducing these impacts. There is an increasing necessity to account for these aspects in the planning, running and end-of-life of scientific experiments and research infrastructure. In the following, the concept for an Open Research Life Cycle Assessment (ORLCA) repository is presented to support this endeavour. It is designed to comply fully with the principles of findability, accessibility, interoperability, and reusability (FAIR).

\vspace{0.4cm}
\keywords{Open Source, Life Cycle Assessment, Database, Repository, Sustainability, Research, Infrastructure, Environment, Greenhouse Gas, GHG, Carbon.}